\documentclass[onecolumn,showpacs,superscriptaddress,amsmath,amssymb]{revtex4-1}
\usepackage[perpage]{footmisc}
\usepackage{epsfig}
\usepackage{dcolumn}
\usepackage{bm}
\usepackage{color}
\usepackage{slashed}

\def\bd{\begin{document}} \def\ed{\end{document}}
\def\bmp{\begin{minipage}} \def\emp{\end{minipage}}
\def\bcc{\begin{center}} \def\ecc{\end{center}}     \def\npg{\newpage}
\def\beq{\begin{equation}} \def\eeq{\end{equation}} \def\hph{\hphantom}
\def\be{\begin{equation}} \def\ee{\end{equation}} \def\r#1{$^{[#1]}$}
\def\n{\noindent} \def\ni{\noindent} \def\pa{\parindent}
\def\hs{\hskip} \def\vs{\vskip} \def\hf{\hfill} \def\ej{\vfill\eject}
\def\cl{\centerline} \def\ob{\obeylines}  \def\ls{\leftskip}
\def\underbar#1{$\setbox0=\hbox{#1} \dp0=1.5pt \mathsurround=0pt
   \underline{\box0}$}   \def\ub{\underbar}    \def\ul{\underline}
\def\f{\left} \def\g{\right} \def\e{{\rm e}} \def\o{\over} \def\d{{\rm d}}
\def\vf{\varphi} \def\pl{\partial} \def\cov{{\rm cov}} \def\ch{{\rm ch}}
\def\la{\langle} \def\ra{\rangle} \def\EE{e$^+$e$^-$} \def\pt{p_{\rm t}}
\def\bitz{\begin{itemize}} \def\eitz{\end{itemize}}
\def\btbl{\begin{tabular}} \def\etbl{\end{tabular}}
\def\btbb{\begin{tabbing}} \def\etbb{\end{tabbing}}
\def\beqar{\begin{eqnarray}} \def\eeqar{\end{eqnarray}}
\def\\{\hfill\break} \def\dit{\item{-}} \def\i{\item}
\def\bbb{\begin{thebibliography}{9} \itemsep=-1mm}
\def\ebb{\end{thebibliography}} \def\bb{\bibitem}
\def\bpic{\begin{picture}(260,240)} \def\epic{\end{picture}}
\def\akgt{\cl{\bf ACKNOWLEDGMENTS}}
\def\fgn{\noindent{\bf\large\bf figure captions}}
\def\lan{\langle}
\def\ran{\rangle}
\def\p{\pi}
\def\ifmath#1{\relax\ifmmode #1\else $#1$\fi}%
\def\rc{\ifmath{{\mathrm{c}}}}
\def\cut{\ifmath{{\mathrm{cut}}}}
\def\rF{\ifmath{{\mathrm{F}}}}
\def\rK{\ifmath{{\mathrm{K}}}}
\def\rp{\ifmath{{\mathrm{p}}}}
\def\rt{\ifmath{{\mathrm{t}}}}
\def\LAB{\ifmath{{\mathrm{LAB}}}}
\def\cut{\ifmath{{\mathrm{cut}}}}
\def\beq{\begin{equation}}
\def\eeq{\end{equation}}

\newcommand{\cinst}[2]{$^{\mathrm{#1}}$~#2\par}
\newcommand{\crefi}[1]{$^{\mathrm{#1}}$}
\newcommand{\crefii}[2]{$^{\mathrm{#1,#2}}$}
\newcommand{\crefiii}[3]{$^{\mathrm{#1,#2,#3}}$}
\newcommand{\HRule}{\rule{0.5\linewidth}{0.5mm}}

\bd

\title{\boldmath Study of $J/\psi\to p\bar{p}$ and $J/\psi\to n\bar{n}$}\\
\author{\small
M.~Ablikim$^{1}$, M.~N.~Achasov$^{5}$, D.~J.~Ambrose$^{40}$, F.~F.~An$^{1}$, Q.~An$^{41}$, Z.~H.~An$^{1}$, J.~Z.~Bai$^{1}$, Y.~Ban$^{27}$, J.~Becker$^{2}$, N.~Berger$^{1}$
% ...
\vspace{0.2cm}\\
(BESIII Collaboration)\\
\vspace{0.2cm} {\it
$^{1}$ Institute of High Energy Physics, Beijing 100049, P. R. China\\
$^{2}$ Bochum Ruhr-University, 44780 Bochum, Germany\\
% ...
$^{5}$ G.I. Budker Institute of Nuclear Physics SB RAS (BINP), Novosibirsk 630090, Russia\\
% ...
$^{27}$ Peking University, Beijing 100871, P. R. China\\
% ...
$^{40}$ University of Rochester, Rochester, New York 14627, USA\\
$^{41}$ University of Science and Technology of China, Hefei 230026, P. R. China\\
% ...
}}

\vspace{0.4cm}


\begin{abstract}
The decays $J/\psi\to p\bar{p}$ and $J/\psi\to n\bar{n}$ have been investigated with a sample of
 225.2 million $J/\psi$ events collected with the BESIII detector at the BEPCII $e^+e^-$ collider. 
The branching fractions are determined to be 
$\mathcal{B}(J/\psi\to p\bar{p})=(2.112\pm0.004\pm0.031)\times10^{-3}$ and
$\mathcal{B}(J/\psi\to n\bar{n})=(2.07\pm0.01\pm0.17)\times10^{-3}$. 
Distributions  of the angle $\theta$ between the proton or anti-neutron and the beam direction
are well described by the form $1+\alpha\cos^2\theta$, and we find $\alpha=0.595\pm0.012\pm0.015$ 
for  $J/\psi\to p\bar{p}$ and $\alpha=0.50\pm0.04\pm0.21$ for $J/\psi\to n\bar{n}$. 
Our branching-fraction results suggest a large phase angle between the strong and electromagnetic 
amplitudes describing the $J/\psi\to N\bar{N}$ decay. 

\end{abstract}


\maketitle

\section{Introduction}\label{intro}

The $J/\psi$ meson is interpreted as a bound state of a charmed quark and a charmed antiquark ($c\bar{c}$). The decay
process $J/\psi\to N\bar{N}$ ($N=p$ or $n$) is an octet-baryon-pair decay mode, and should be a good laboratory for testing 
perturbative QCD (pQCD) because the three gluons in the OZI-violating strong decay correspond to the three $q\bar{q}$ pairs
that form the final-state nucleons. The ratio of the branching fractions for the $p\bar{p}$ and $n\bar{n}$ final states provides information about the phase angle between the strong and the electromagnetic (EM) amplitudes governing the decay~\cite{ref:theory01, Yuan:2003hj, Wang:2003hy}.  Because  the initial-state isospin is 0, the strong-decay amplitudes for the $p\bar{p}$ and $n\bar{n}$ final states must be equal.  The $J/\psi\to p\bar{p}$  and  $J/\psi\to n\bar{n}$ EM decays are expected to have amplitudes that are of about the same magnitude, but with opposite signs, like the magnetic moments (as discussed in Section~\ref{sec-summary}).  Because the EM decays of $J/\psi$ to $p{\bar p}$ and $n{\bar n}$ behave the same as non-resonant production of those final states, the magnitude of the EM decay amplitude of $J/\psi$ can be estimated from the cross section for continuum production $e^+e^- \to p{\bar p}$. If the strong and EM amplitudes are almost real, and therefore in phase, as predicted by pQCD~\cite{ref:theory01, Yuan:2003hj, Wang:2003hy, ref:theory3, ref:theory4}, then interference would lead to a branching fraction for $J/\psi\to n\bar{n}$ about one-half as large as that for $J/\psi\to p\bar{p}$.  Conversely, if the strong and EM amplitudes are orthogonal, then the strong decay dominates and the branching fractions are expected to be equal. In previous experiments, $J/\psi\to p\bar{p}$ has been measured with good precision, while  $J/\psi\to n\bar{n}$ has been measured with quite a large uncertainty ~\cite{ref:pdg, ref:finice}.  They appear to be equal within errors, at odds with the pQCD expectation.

The angular distribution for $J/\psi\to N\bar{N}$  can be written as a function of the angle $\theta$ between 
the nucleon or antinucleon direction and the beam as follows:

$$\frac{d N}{d\cos\theta} = A (1+ \alpha\cos^2\theta),$$

\noindent where $A$ is an overall normalization. These angular distributions reflect details of the baryon structure 
and have the potential to distinguish among different theoretical models~\cite{ref:theory01, Yuan:2003hj, Wang:2003hy, ref:theory3,ref:theory4}.

In this paper, we report new studies of the process $J/\psi\to N\bar{N}$ made with the BESIII detector at the BEPCII 
electron-positron storage ring~\cite{ref:bes3,ref:bes3physics}. With the world's largest sample of $J/\psi$ decays, 
we obtain improved measurements for the $J/\psi\to p\bar{p}$ and $J/\psi\to n\bar{n}$ branching fractions and angular 
distributions.

\section{BEPCII and BESIII}\label{bepc2bes3}

BEPCII is a two-ring $e^+e^-$ collider designed for a peak luminosity of $10^{33}$ cm$^{-2}s^{-1}$ at a beam current 
of 0.93~A.  The cylindrical core of the BESIII detector consists of a helium-gas-based drift chamber~(MDC) for 
charged-particle tracking and particle identification by $dE/dx$, a plastic scintillator time-of-flight system~(TOF) 
for additional particle identification, and a 6240-crystal CsI(Tl) Electromagnetic Calorimeter~(EMC) for electron 
identification and photon detection.  These components are all enclosed in a superconducting solenoidal magnet providing
a 1.0-T magnetic field.  The solenoid is supported by an octagonal  flux-return yoke with resistive-plate-counter muon 
detector modules (MU) interleaved with steel.  The geometrical acceptance for charged tracks and photons is $93\%$ of
 $4\pi$, and the resolutions for charged-track momentum and photon energy at 1~GeV are $0.5\%$ and $2.5\%$, respectively.  
More details on the features and capabilities of BESIII are provided in Ref.~\cite{ref:bes3}.

\section{Data sample}\label{datasample}
Our data sample consists of 225.2 million $e^+ e^-\to J/\psi$ events collected during 2009.  
The estimated uncertainty in the number of events is $\pm 1.3\%$~\cite{ref:jpsitotnumber}.  A 
{\tt GEANT4}-based~\cite{Agostinelli:2002hh,Allison:2006ve} detector simulation is used to produce Monte Carlo 
(MC) samples for signal and background processes that are generated with specialized models that have
been packaged and customized for BESIII~\cite{ref:bes3gen}. {\tt EvtGen}~\cite{ref:evtgen} is used to study phase-space signal events for $J/\psi\to p\bar{p}$ and for exclusive backgrounds in $J/\psi$ decays.  
{\tt BABAYAGA}~\cite{ref:babayaga} is used  to generate Bhabha and $\gamma\gamma$ events as possible EM 
backgrounds.  A large inclusive sample (200 million events) is used to simulate hadronic background processes.  
The $J/\psi$ resonance is generated by {\tt KKMC}~\cite{ref:kkmc}.  Known $J/\psi$ decay modes are generated with 
{\tt EvtGen}, using branching fractions set to world-average values~\cite{ref:pdg}.  The remaining $J/\psi$ decay 
modes are generated by {\tt LUNDCHARM}~\cite{ref:bes3gen}, which is based on {\tt JETSET}~\cite{Sjostrand:1993yb} 
and tuned for the charm-energy region.  The decays $J/\psi\to p\bar{p}$ and $J/\psi\to n\bar{n}$ are excluded from 
this sample.  

\section{General event selection}\label{data sample}

Charged tracks in BESIII are reconstructed from MDC hits.  To
optimize the momentum measurement, we select tracks in the polar
angle range $|\cos\theta|~<~0.93$ and require that they pass within
$\pm 10$~cm of the interaction point in the beam direction and
within $\pm 1$~cm in the plane perpendicular to the beam. 

Electromagnetic showers are reconstructed by clustering EMC crystal
energies.  Efficiency and energy resolution are improved by
including energy deposits in nearby TOF counters.  Showers used in
selecting photons and in $\pi^0$ reconstruction must
satisfy fiducial and shower-quality requirements.  Showers in the
barrel region ($|\cos\theta|<0.8$) must have a minimum energy of
25~MeV, while those in the endcaps ($0.86 < |\cos\theta| < 0.92$)
must have at least 50~MeV.  Showers in the region between the
barrel and endcap are poorly reconstructed and are excluded. To 
eliminate showers from charged particles, a photon must
be separated by at least 10$^\circ$ from any charged track.  EMC
timing requirements suppress electronic noise and energy deposits 
unrelated to the event.

\section{Analysis of $J/\psi\to p\bar{p}$}\label{jpsipp}

\subsection{\bf Event Selection}\label{sec-selppbar}

Events with exactly two good charged tracks in the polar angle range $|\cos\theta| < 0.8$ are selected. We exclude 
the two endcap regions to reduce systematic uncertainties in tracking and particle identification. By using a 
loose particle-identification requirement for the positive track (probability of the $p$ hypothesis greater than 
the probabilities for the $\pi^+$ and $K^+$ hypotheses), and by requiring no particle identification for the negative 
track, the efficiency is maximized and the systematic uncertainty is minimized.  A vertex fit is performed to the two 
selected tracks to improve the momentum resolution, and the angle between the $p$ and $\bar{p}$ is required to be 
greater than $178^\circ$.  Finally, for both tracks, the measured momentum magnitude must be within 30~MeV/$c$ 
($\sim 3\sigma$) of the expected value of 1.232~GeV/$c$.  Fig.~\ref{fig-ppbardatamc} shows comparisons between 
data and MC for the angle between the $p$ and $\bar{p}$ and for their momenta. 

This selection results in a signal of $N=314651 \pm 561$ candidate events. Fig.~\ref{fig-ppbardatamc}~(b) and (c)
show the $p$ and ${\bar p}$ momentum distributions for these events, along with the expected distributions for
a pure MC $J/\psi\to p\bar{p}$ signal.  Backgrounds overall are very small, and appear to be negligible in the 
accepted $p$ and ${\bar p}$ momentum range.  Three independent procedures are used to estimate this background: 
inclusive $J/\psi$ MC, exclusive MC of potential background processes (Bhabha events and $J/\psi$ decays to $e^+e^-$, 
$\mu^+\mu^-$, $K^+K^-$, $\gamma p\bar{p}$, $\pi^0 p\bar{p}$, and $\gamma\eta_c$ with $\eta_c\to p\bar{p}$), and a 
sideband technique. The estimates range from 0.02\% to $0.2\%$ of the signal.  We apply no subtraction and take the 
largest of the estimates (sideband) as a systematic uncertainty in the final result.  The raw distribution of 
$\cos \theta$ for the protons in the selected signal events is given in Fig.~\ref{fig-ppbarangleside}.

\begin{figure}[htbp]
\begin{center}
\includegraphics[width=16cm]{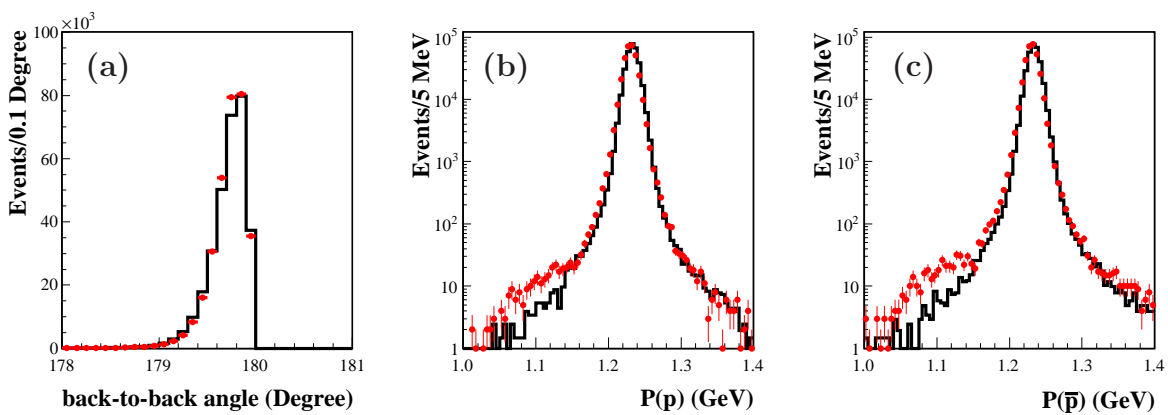}
\put(-425,140){\bf \large~(a)}
\put(-275,140){\bf \large~(b)}
\put(-120,140){\bf \large~(c)}
\caption{Comparisons between data (points) and MC (histograms) for properties of the $p$ and $\bar{p}$  tracks for 
selected $J/\psi\to p\bar{p}$ signal events:~(a)  angle between the $p$ and $\bar{p}$, (b)~$p$ momentum, 
and (c)~$\bar{p}$ momentum. } 
\label{fig-ppbardatamc}
\end{center}
\end{figure}

\begin{figure}[htbp]
\begin{center}
\includegraphics[width=8cm]{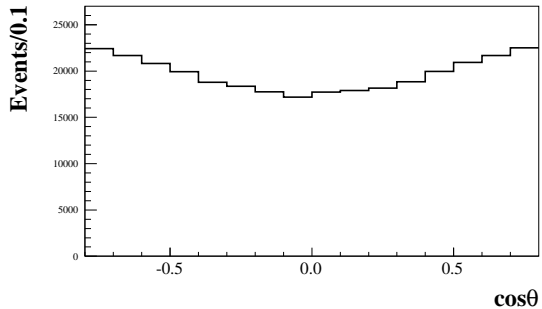}
\caption{Angular distribution of the selected $J/\psi\to p\bar{p}$ candidates.} \label{fig-ppbarangleside}
\end{center}
\end{figure}

\bigskip
\subsection{\bf Efficiency Correction}\label{sec-effcor}
\bigskip

To measure the $J/\psi \to p\bar{p}$ branching fraction and angular distribution, it is necessary to correct for the 
selection efficiency, which is dominated by track reconstruction and selection and by particle-identification 
efficiency.  We use signal MC to obtain the efficiency, but use data to correct for imperfections in the simulation, 
thereby reducing the systematic uncertainty in the correction.  We measure differences between the data and MC 
separately for the efficiencies of tracking and particle identification, leaving the other selection cuts in place
or tightening them for cleaner selection.  The correlation between the corrections in the tracking and 
particle-identification efficiencies has been shown in MC studies to be small, so we combine them into a single 
correction function that is applied to the MC-determined efficiency.  Because the tracking and TOF response depend 
on the track direction, the efficiency correction is determined in bins of $\cos\theta$.

We divide the full angular range ($|\cos \theta| < 0.8$) into 16 equal bins and for each bin compute the 
efficiency for the successful reconstruction of the ${\bar p}$ or $p$ track as follows:

$$\epsilon_{trk}=\frac{N_{2}}{N_{1}+N_{2}},$$

\noindent where $N_{1}$ ($N_{2}$) is the number of $J/\psi\to p\bar{p}$ events with 1 (2) 
good charged track(s) detected. For $N_{1}$, we require only one good charged track which is identified 
as a $p$ or $\bar{p}$. Note that in this case, unlike the $J/\psi \to p {\bar p}$ selection, we can apply 
particle identification to the $\bar{p}$ selection to improve purity, since any inconsistency between data 
and MC would cancel in the efficiency.  Fig.~\ref{fig-calibtrk} shows the data/MC comparison for the tracking 
efficiencies and the computed correction factor $\epsilon^{data}_{trk}/\epsilon^{MC}_{trk}$  for each 
$\cos\theta$ bin for $p$ and $\bar{p}$. 

\newpage
\begin{figure*}[htbp]
\begin{center}
\includegraphics[width=16cm]{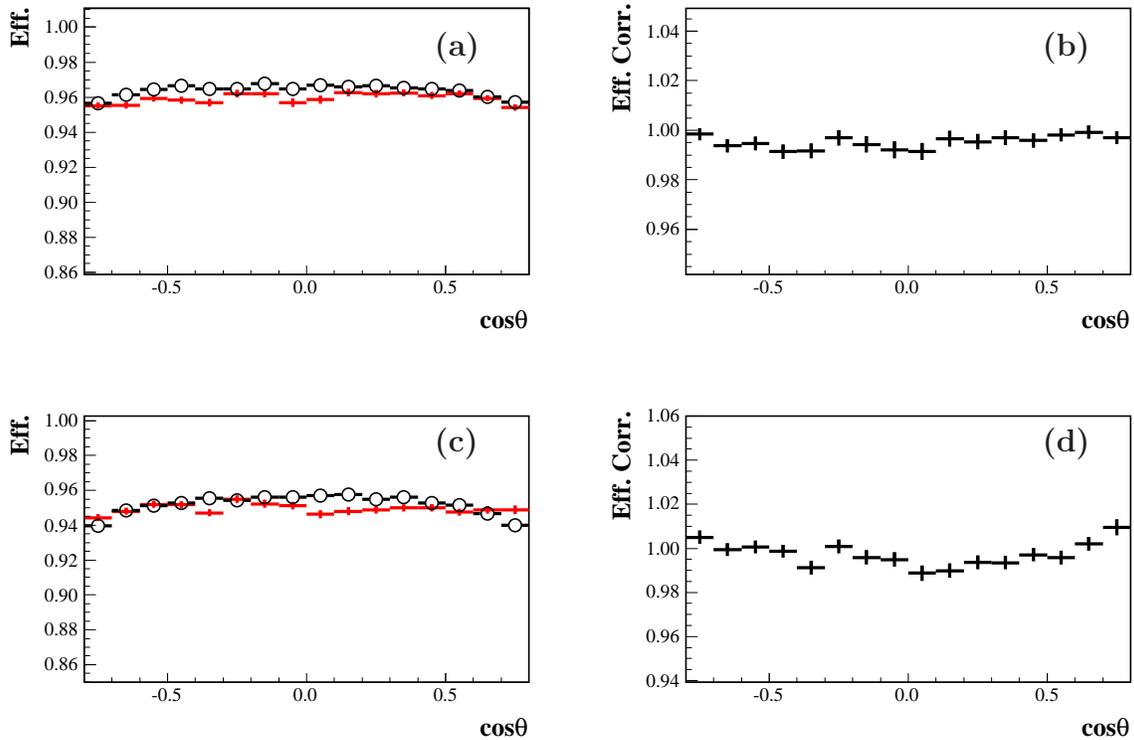}
\put(-290,270){\bf \large~(a)}
\put(-60,270){\bf \large~(b)}
\put(-290,120){\bf \large~(c)}
\put(-60,120){\bf \large~(d)}
\caption{(a) The proton tracking efficiency for data (points) and MC (circles), and (b) the correction 
$\epsilon^{data}_{trk}/\epsilon^{MC}_{trk}$; (c) and (d) show the same for antiprotons.}\label{fig-calibtrk}
\end{center}
\end{figure*}

We can similarly measure the particle-identification efficiency for the $p$ in each $\cos\theta$ bin, 
considering only $J/\psi\to p\bar{p}$ events in which there are two good charged tracks, with the 
negatively-charged track identified as an antiproton.  We define the efficiency as follows:

$$\epsilon_{pid}=\frac{N_p}{N_p+N_{\slashed p}},$$

\noindent where $N_p$ is the number of selected events in which the proton has been successfully identified and
$N_{\slashed p}$ is the number of events without the proton identified.  To select a more pure sample we 
tighten the selection on the $p$ and $\bar{p}$ momenta to be within 20~MeV/$c$ ($\sim 2\sigma$) of the
expected value.  Figure~\ref{fig-calibpid} shows the data/MC comparison for the proton particle identification 
efficiency and the resulting correction factor $\epsilon^{data}_{pid}/\epsilon^{MC}_{pid}$.

\begin{figure*}[htbp]
\begin{center}
\includegraphics[width=16cm]{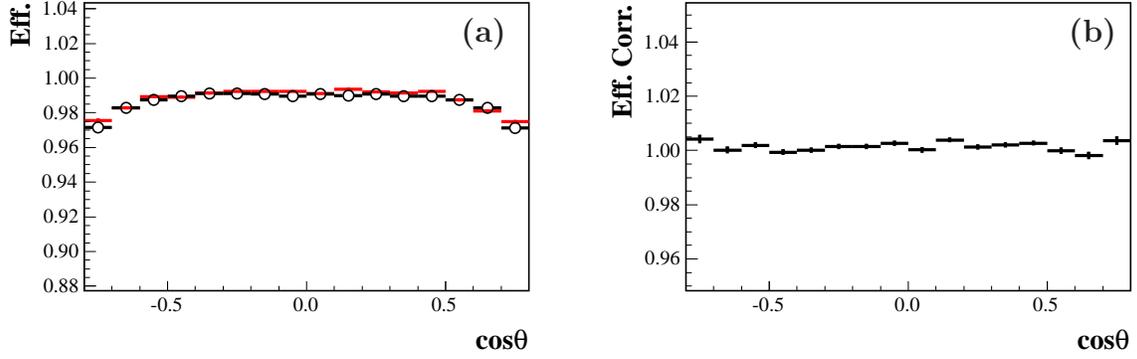}
\put(-280,128){\bf \large~(a)}
\put(-50,128){\bf \large~(b)}
\caption{(a)~$p$ particle-identification efficiency for data (points) and MC (circles), and (b) the computed 
efficiency correction factor $\epsilon^{data}_{pid}/\epsilon^{MC}_{pid}$.} \label{fig-calibpid}
\end{center}
\end{figure*}

In each $\cos\theta$ bin, the corrected MC-determined efficiency to be applied to data is computed with the
following formula: 

$$\epsilon = \epsilon^{MC} \times \frac{\epsilon_{ptrk}^{data}}{\epsilon_{ptrk}^{MC}} \times\frac{\epsilon_{ppid}^{data}}{\epsilon_{ppid}^{MC}}\times \frac{\epsilon_{\bar{p}trk}^{data}}{\epsilon_{\bar{p}trk}^{MC}}.$$

To diminish the effect of bin-to-bin scatter due to statistical fluctuations, we fit the corrected efficiency
as a function of $\cos \theta$ with a fifth-order polynomial, as shown in Fig.~\ref{fig-toteff}. 

\begin{figure*}[htbp]
\begin{center}
\includegraphics[width=16cm]{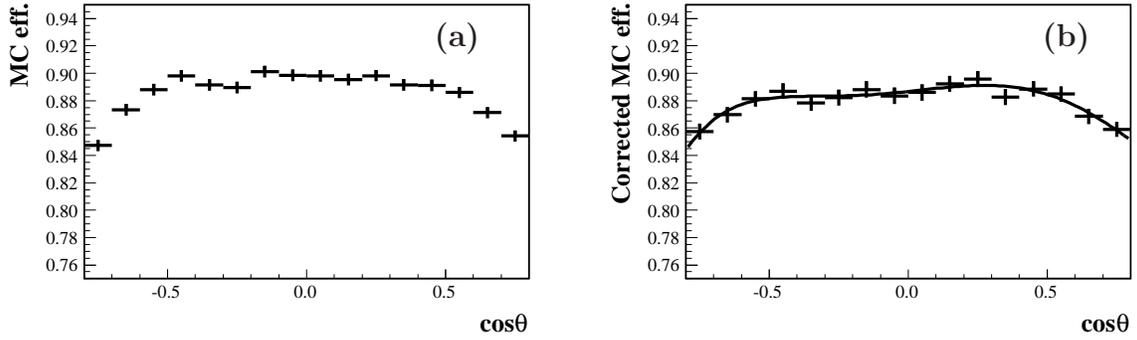}
\put(-290,120){\bf \large~(a)}
\put(-60,120){\bf \large~(b)}
\caption{$J/\psi\to p\bar{p}$ selection efficiency as a function of $\cos\theta$~(a)~before correction, 
and (b) after correction.  The line shows the smoothed efficiency obtained by fitting the data to a 
fifth-order polynomial.} \label{fig-toteff}
\end{center}
\end{figure*}

\subsection{\bf Angular Distribution and Branching Fraction}

We fit the measured angular distribution of the proton from $J/\psi \to p {\bar p}$ to the function 
$A(1+\alpha\cos^2\theta)\epsilon(\cos\theta)$, where $A$ is the overall normalization and 
$\epsilon(\cos\theta)$ is the corrected MC-determined efficiency function (Sect.~\ref{sec-effcor}).  
The angular distribution and the fit are shown in Fig.~\ref{fig-fitangular}. The $\chi^2$ for the fit 
is 16, with 14 degrees of freedom, and the value determined for the angular-distribution parameter 
is $\alpha = 0.595\pm0.012$, where the error is statistical only.

\begin{figure}[htbp]
\begin{center}
\includegraphics[width=8cm]{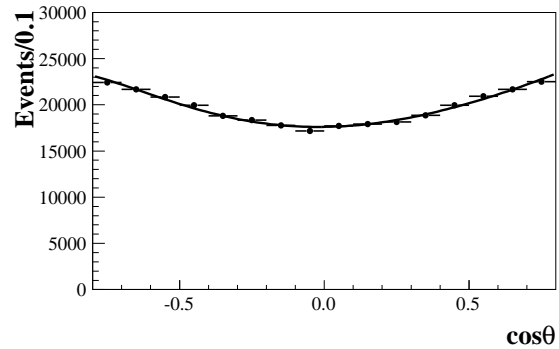}
\caption{The points represent the measured distribution of $\cos \theta$ for the $p$ in $J/\psi\to p\bar{p}$ 
candidate events, with error bars that are the quadratic sums of the statistical and efficiency uncertainties. 
The line represents the fit of the distribution to the functional form given in the text, and is used to determine 
the normalization and the angular-distribution parameter $\alpha$.} \label{fig-fitangular}
\end{center}
\end{figure}

The raw yield of $J/\psi \to p {\bar p}$ events obtained by counting protons in the angular range 
$\cos\theta=[-0.8,0.8]$ is $N(-0.8,0.8) = 314651 \pm 561$.  The efficiency-corrected yield obtained by
fitting the $\cos \theta$ distribution over this range is $N_{cor}(-0.8,0.8) = 357786 \pm638$. The fitted value 
of $\alpha$ is used to determine the total number of $J/\psi \to n {\bar n}$ events in the full angular range 
of $\cos\theta=[-1.0,1.0]$ as follows:

$$N_{cor}(-1.0,1.0) = N_{cor}(-0.8,0.8) \times \frac{\int_{-1.0}^{1.0}(1+\alpha\cos^2\theta)d\theta}{\int_{-0.8}^{0.8}(1+\alpha\cos^2\theta)d\theta}=475567 \pm848.$$

Combining this final yield with the number of $J/\psi$ events in our sample ($(2.252 \pm 0.029) \times 10^8)$, 
we find the branching fraction to be

$$\mathcal{B}(J/\psi\to p\bar{p})=(2.112\pm0.004)\times10^{-3},$$

\noindent where the error is statistical only.

\subsection{\bf Systematic Errors and Results}
To determine the uncertainty in the efficiency correction, we use toy MC experiments to 
obtain distributions in the branching fraction and $\alpha$ that reflect the statistical errors of the bin-by-bin 
efficiency values. We perform this study by varying each bin randomly according
to a normal distribution for each MC experiment, redoing the polynomial fit and then remeasuring the efficiency-corrected yield. The results have normal distributions and they are fitted with Gaussian functions to estimate the
associated uncertainties in the branching fraction and $\alpha$, which are found to be 
$4.69\times10^{-6}$  and 0.011, respectively.  

The full magnitude of the $p\bar{p}$ momentum sideband background estimate ($0.2\%$) is taken to be 
the uncertainty in the branching fraction due to the background correction.  We fit the 
sideband-subtracted angular distribution and determine a new value for the angular parameter $\alpha$,
taking the change relative to the standard result (0.004) as the systematic error.  

To estimate the systematic error due to the detector angular resolution, we perform a study with the 
signal MC.  The ``true'' generated proton $\cos\theta$ is fitted before and after smearing with a 
MC-derived angular resolution function.  The differences in the fitted $\alpha$ values
(0.010) and in the branching fractions ($4\times10^{-6}$) are taken as the systematic uncertainties
from this source.
  
The branching fraction also incurs two systematic uncertainties that do not affect $\alpha$.  A small 
systematic uncertainty enters due to the correction for the $|\cos \theta|<0.8$ requirement, which 
depends on the determined value of $\alpha$ and its error.  The dominant uncertainty in the 
branching fraction is due to the estimated 1.3\% error in the number of $J/\psi$ events in our
sample~\cite{ref:jpsitotnumber}.  

To study the effect from continuum production, we write the total cross section $\sigma_{p\bar{p}}$ as 

$$\sigma_{p\bar{p}} = |\sqrt{\sigma_{p\bar{p}}^{cont.}} + \frac{\sqrt{12\pi\Gamma_{ee}\Gamma_{tot}}}{s-m^2+im\Gamma_{tot}}(E_{p}+e^{i\phi}S)|^2,$$ 

\noindent where $E_{p}$ and $S$ are the EM and strong amplitudes of $J/\psi\to p\bar{p}$ and $\phi$ is the phase angle between them. $\sigma_{p\bar{p}}^{cont.}$ is the $p\bar{p}$ cross section contributed by the continuum under the $J/\psi$ peak. These values are taken from the calculation in Sect.~\ref{sec-summary}. The difference with and without $\sigma_{p\bar{p}}^{cont.}$ is assigned as the systematic error.

Finally, we change $\alpha$  by $\pm 1$~$\sigma$ (includes
systematic error) and reevaluate the branching fraction to estimate the systematic error in the branching fraction 
measurement. 

Table~\ref{tab-ppbarsys} provides a summary of all identified
sources of systematic uncertainty, which are assumed to be uncorrelated, and their quadrature sum.   
The final results for our $J/\psi\to p\bar{p}$ measurements are as follows:

$$\alpha=0.595\pm0.012\pm0.015,~{\rm and}$$

$$\mathcal{B}(J/\psi\to p\bar{p})=(2.112\pm0.004\pm0.031)\times10^{-3}.$$

\noindent The branching fraction measurement is consistent with the previous world 
average~\cite{ref:pdg} and improves the overall precision by about a factor of 2.5. The value of 
$\alpha$ is also consistent with previous experiments (Table~\ref{tab-ppbaralpha}) and is improved 
significantly. 

\begin{table*}[htbp]
\begin{center}
\caption{Systematic errors for $J/\psi\to p\bar{p}$.} \label{tab-ppbarsys}
\vspace{0.2cm}{
\begin{tabular}{|c|c|c|}\hline\hline
Sources & Effect on $\alpha$ & Effect on  $\mathcal{B}$($10^{-3}$)\\
\hline\hline
Efficiency Correction & 0.011 & 0.005\\
Background & 0.004  & 0.002\\
$\cos\theta$ Resolution & 0.010 & 0.004\\
$\alpha$ Value & $-$ & 0.004\\
Number of $J/\psi$ & $-$ & 0.026\\
continuum & $-$ & 0.015\\
\hline
Total & 0.015 & 0.031 \\
\hline\hline
\end{tabular}
}
\end{center}
\end{table*}

\begin{table*}[htbp]
\begin{center}
\caption{ Previous measurements of $\alpha$ in $J/\psi\to p\bar{p}$.} \label{tab-ppbaralpha}
\vspace{0.2cm}{
\begin{tabular}{|c|c|}\hline
Collaboration & $\alpha$\\
\hline\hline
Mark1 \cite{Peruzzi:1977pb} & $1.45\pm0.56$\\
Mark2 \cite{Eaton:1983kb} & $0.61\pm0.23$\\
Mark3 \cite{Mark3}& $0.58\pm0.14$\\
DASP~\cite{DSAP} & $1.70\pm1.70$\\
DM2~\cite{Pallin:1987py} & $0.62\pm0.11$\\
BESII~\cite{Bai:2004jg} & $0.676\pm0.055$\\
\hline
\end{tabular}
}
\end{center}
\end{table*}

\section{\bf Analysis of $J/\psi\to n\bar{n}$}

\subsection{\bf Event Selection}\label{sec-selnnbar}

We search for $J/\psi\to n\bar{n}$ candidates by selecting events that have no good charged tracks 
originating in the interaction region. The antineutron annihilation ``star'' in the EMC provides
a signature for these events that is much more identifiable than the hadronic shower produced by a 
neutron.  We therefore first select events with showers characteristic of $\bar{n}$ interactions, 
and then search in these events for energy deposited by $n$ hadronic interactions on the opposite side 
of the detector. 

The most energetic shower in the event is assigned to be the $\bar{n}$ candidate and is required to 
have an energy in the range $0.6-2.0$~GeV.  To optimize the discrimination against backgrounds, 
we apply a fiducial cut of $|\cos\theta|<0.8$ to the ${\bar n}$ candidate.  This ensures that the 
${\bar n}$ energy is fully contained in the EMC for most signal events.   To suppress photon 
backgrounds, we impose a requirement on the second moment of the candidate shower, defined as 
$S=\Sigma_{i}E_{i}r_{i}^{2}/\Sigma_{i}E_{i}$, where $E_{i}$ is the energy deposited in the $i^{th}$ 
crystal of the shower and $r_{i}$ is the distance from the center of that crystal to the center of 
the shower.  To be accepted, the $\bar{n}$ candidate must satisfy $S>20$~cm$^2$.  To further exploit the 
distinctive $\bar{n}$ shower topology, we require the number of EMC hits in a $50^\circ$ cone around 
the $\bar{n}$ candidate shower direction to be greater than 40.  

Events with accepted $\bar{n}$ candidates are searched for EMC showers on the opposite side of the 
detector that are consistent with being the neutron in a $J/\psi\to n\bar{n}$ decay.  The energy of 
this shower must be between $0.06$ and $0.6$~GeV, a range found to be characteristic of the EMC 
neutron response in MC studies.  If multiple showers are present, the one that is most back-to-back 
with respect to the $\bar{n}$ candidate is selected.  To further suppress backgrounds from 
all-neutral $J/\psi$ decays, continuum production and EM processes, we require $E_{extra}=0$, where 
$E_{extra}$ is the total deposited energy in the EMC, excluding that of the $n$ shower and any 
additional energy in the $50^{\circ}$ cone.

The expected signal for $J/\psi\to n\bar{n}$ is an enhancement near $180^{\circ}$ in the distribution 
of the angle between the $n$ shower and the direction of the $\bar{n}$.  The distributions of this 
angle and of the cosine of the polar angle of the $\bar{n}$ shower ($\cos\theta$) for selected 
candidates are shown in Fig.~\ref{fig-nnbarpasssel}.  The enhancement near $180^{\circ}$ in 
the distribution of the angle between the $n$ and $\bar{n}$ constitutes the $J/\psi\to n\bar{n}$ 
signal.  Since there is nonnegligible background, the number of $J/\psi\to n\bar{n}$ events must be 
determined by fitting.  Distributions of the angle between the $n$ and $\bar{n}$ are constructed in
bins of $\cos \theta$ and fitted with signal and background functions.

Data-driven methods are used to determine the efficiency and signal shapes for $J/\psi\to n\bar{n}$. 
We select $J/\psi\to p (\bar{n})\pi^-$ and charge-conjugate (c.c.) events in data to obtain $n$ 
and $\bar{n}$ samples to evaluate the selection efficiency.   We use the $p$ and $\bar{p}$  in $J/\psi\to p\bar{p}$ 
events that have been selected using information from just the MDC to get unbiased information on the
shape and efficiency of the $n$ and $\bar{n}$ response in the EMC, since antiproton and antineutron hadronic interactions are similar. 

Generic $J/\psi$ MC is used to assess the background.  Fig.~\ref{fig-nnbarpasssel}~(a) shows that 
there is no peaking in the distribution of the angle between the $n$ and $\bar{n}$ for this background.  
We also consider possible exclusive background channels: $J/\psi\to \pi^0 n\bar{n}$, 
$J/\psi\to \gamma n\bar{n}$, $e^+e^-\to\gamma\gamma$, $J/\psi\to\bar{\Sigma}^+\Sigma^-$, 
 $J/\psi\to \Sigma^+\bar{\Sigma}^-$, $J/\psi\to p\bar{p}$, and $J/\psi\to\gamma\eta_c$ ($\eta_c\to n\bar{n}$).  
None of these potential background sources exhibits peaking in the distribution of the angle between the 
$n$ and $\bar{n}$.

\begin{figure*}[htbp]
\begin{center}
\includegraphics[width=16cm]{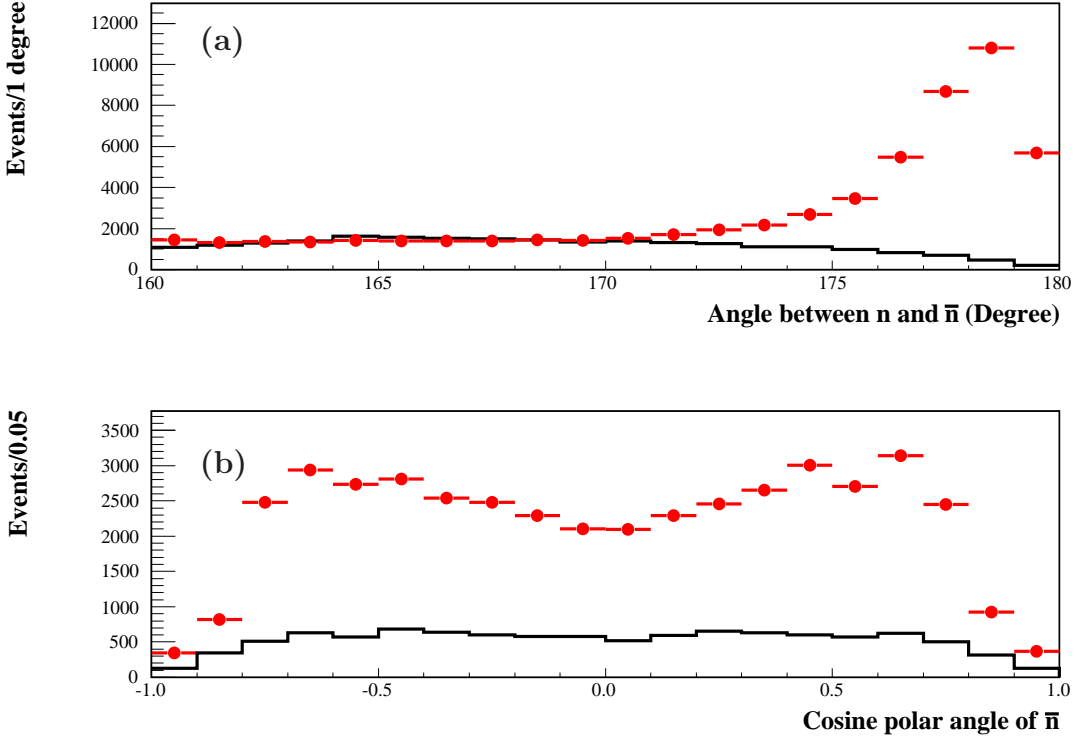}
\put(-370,270){\bf \large~(a)}
\put(-370,110){\bf \large~(b)}
\caption{Distributions for $J/\psi\to n\bar{n}$ candidate events summed over all $\cos\theta$ bins 
(points) and background from inclusive MC (solid lines):~(a)~angle between the $n$ and $\bar{n}$, 
and~(b)~$\cos \theta$ for the $\bar{n}$ shower.} \label{fig-nnbarpasssel}
\end{center}
\end{figure*}

\subsection{\bf Efficiency Determination}\label{sec-effnnbar}

We use specially chosen event samples from data to determine the efficiencies for each requirement 
in the $J/\psi\to n\bar{n}$ selection.  The overall efficiency is then computed bin-by-bin in 
$\cos \theta$ as the product of these components and is applied as a correction in obtaining the 
angular distribution and branching fraction. 

We select $J/\psi\to p\bar{n}\pi^-$ events to study the efficiency of the $\bar{n}$ selection.  
Events with exactly two good charged tracks identified as $p$ and $\pi^-$  are selected.  
Information from the TOF detector and $dE/dx$ information from the MDC are combined to do the 
particle identification.  The $p$ and $\pi^-$ are required to have a missing mass within 
$30~$MeV of the nominal $\bar{n}$ mass.  The missing momentum of the $p$ and $\pi^-$ is 
required to be in the range $1.1-1.2$~GeV/$c$ to ensure a sample that is as similar as 
possible to the $\bar{n}$ in $J/\psi\to n\bar{n}$ (momentum 1.232~GeV/$c$).  The number 
of events passing the above selection gives $N^{exp}$, the expected $\bar{n}$ yield.  The 
number of $\bar{n}$ candidates selected from these events (criteria defined in 
Sect.~\ref{sec-selnnbar}) that match the missing momenta of the accompanying $p$ and $\pi^-$ 
within $10^\circ$ gives the observed yield $N^{obs}$.  The efficiency for $\bar{n}$ selection 
is $\epsilon^{data}_{\bar{n}}=N^{obs}/N^{exp}$. 

To validate this procedure and ensure consistency between the $\bar{n}$ in  
$J/\psi\to p\bar{n}\pi^-$ and that in the signal process $J/\psi\to n\bar{n}$, we select 
higher-purity $\bar{n}$ candidates in $J/\psi\to n\bar{n}$ ($J/\psi\to p\bar{n} \pi^-$) with 
a stringent cut of $177^\circ$ on the angle between the $n$ and $\bar{n}$.  (For 
$J/\psi\to p\bar{n}\pi^-$ the cut is on the angle between the $\bar{n}$ and the missing 
momentum of  the $p$ and $\pi^-$.)  Comparisons of the selection variables (energy deposit in 
EMC, number of EMC hits in a $50^\circ$ cone about the $\bar{n}$ shower, and the shower second 
moment) for these two $\bar{n}$ samples are shown in Fig.~\ref{fig-nbarpnpi}.   Each distribution 
is plotted after the cuts on the other variables have been imposed.  There is good agreement, 
verifying that the $\bar{n}$ in $J/\psi\to p(\bar{n}) \pi^-$ is a good match to the $\bar{n}$ in 
the signal process $J/\psi\to n\bar{n}$, and that this process provides a reliable efficiency 
correction.

We apply the same technique to calculate the efficiency for selecting the neutron 
($\epsilon^{data}_{n}$), in this case using a sample of $J/\psi\to \bar{p}( n )\pi^+$ selected 
from data.  A comparison of the distribution of the EMC energy for neutrons from 
$J/\psi\to \bar{p}( n )\pi^+$ with that from $J/\psi\to n\bar{n}$ is shown in Fig.~\ref{fig-n0pnpi}.  
In this case the momentum difference between the two $n$ samples results in a greater difference in 
the EMC energy than was observed in the $\bar{n}$ case.  This disagreement is a source of 
systematic error, which we try to minimize by the use of the very loose energy cut on the $n$ 
shower.

\begin{figure}[htbp]
\begin{center}
\includegraphics[width=16cm]{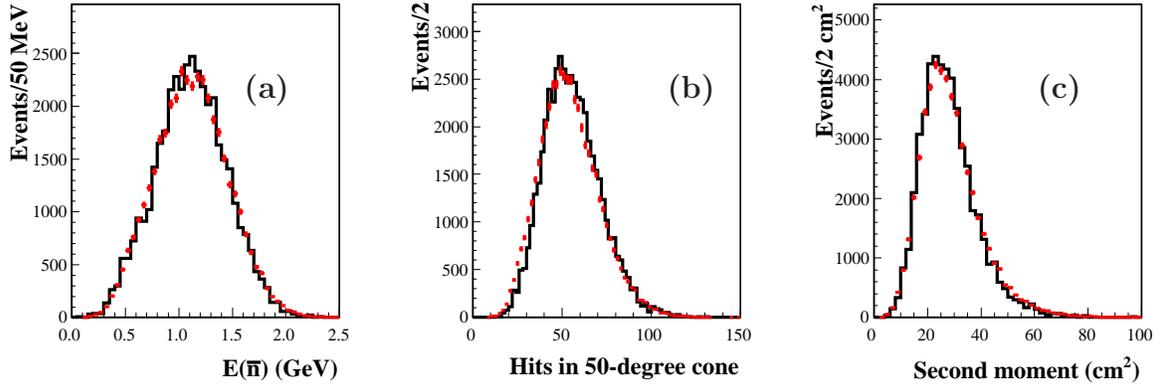}
\put(-360,120){\bf \large~(a)}
\put(-200,120){\bf \large~(b)}
\put(-60,120){\bf \large~(c)}
\caption{Comparisons of distributions of selection variables for $\bar{n}$ from $J/\psi\to n\bar{n}$ 
(solid line) with those from $J/\psi\to p \bar{n}\pi^-$ (points): (a) deposited energy in the 
EMC,~(b)~the number of EMC hits in the $50^\circ$ cone around the $\bar{n}$ shower, and (c)~the 
second moment of the EMC energy deposit.} \label{fig-nbarpnpi}
\end{center}
\end{figure}

\begin{figure}[htbp]
\begin{center}
\includegraphics[width=8cm]{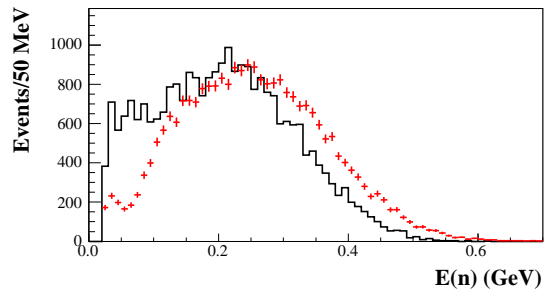}
\caption{Comparison of the distribution of the deposited energy in the EMC for $n$ showers from 
$J/\psi\to n\bar{n}$ (points) with that from $J/\psi\to \bar{p} n\pi^+$ (solid line).} 
\label{fig-n0pnpi}
\end{center}
\end{figure}

For the $E_{extra}$ cut, we use $J/\psi\to p\bar{p}$ to obtain the efficiency 
($\epsilon^{data}_{E_{extra}}$). The requirements are identical to those described in 
Sect.~\ref{sec-selppbar}.  Our selection of $J/\psi\to p\bar{p}$ does not depend on information 
from the calorimeter, so the behavior of $p\bar{p}$ in the EMC can be used to verify the efficiency 
of the $E_{extra}$ cut for $J/\psi\to n\bar{n}$.  Figure~\ref{fig-emissppbar} shows 
the comparison of the $E_{extra}$ distributions for $J/\psi\to p\bar{p}$ and $J/\psi\to n\bar{n}$.  
We require the angle between the $n$ and $\bar{n}$ to be greater than $177^\circ$ to suppress background 
for this comparison.  We find that the proportion of $E_{extra}=0$ events in $J/\psi\to p\bar{p}$  and 
$J/\psi\to n\bar{n}$ agree well. The ratio of $J/\psi\to p\bar{p}$ events with or without the 
requirement $E_{extra}=0$ is calculated as the efficiency of the $E_{extra}$ cut.

\begin{figure}[htbp]
\begin{center}
\includegraphics[width=8cm]{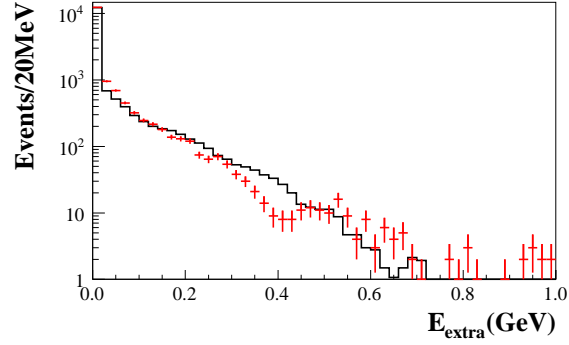}
\caption{Comparison of the distribution of $E_{extra}$ in $J/\psi\to n\bar{n}$ (points) with that 
from $J/\psi\to p\bar{p}$ (solid line).} \label{fig-emissppbar}
\end{center}
\end{figure}

Finally, we determine the overall efficiency from the product of the three component efficiencies 
described above:

$$\epsilon=\epsilon^{data}_{\bar{n}}\epsilon^{data}_{n}\epsilon^{data}_{E_{extra}}.$$

\noindent To facilitiate measurement of the angular distribution as well as the yield, the 
product efficiency is determined in 16 bins in $\cos \theta$ (cosine of the $\bar{n}$ polar 
angle) from -0.8 to 0.8.  Figure~\ref{fig-effnnbar} shows the efficiency for each cut and the product 
as a function of $\cos \theta$.  The loss  of $\bar{n}$ efficiency near $\cos \theta = \pm0.8$ is 
caused by the requirement on the number of EMC hits in a $50^\circ$ cone around the 
shower.  To smooth the bin-to-bin statistical fluctuations in the efficiency 
correction, we fit with a fifth-order polynomial (Fig.~\ref{fig-effnnbar}(d)).

\begin{figure}[htbp]
\begin{center}
\includegraphics[width=16cm]{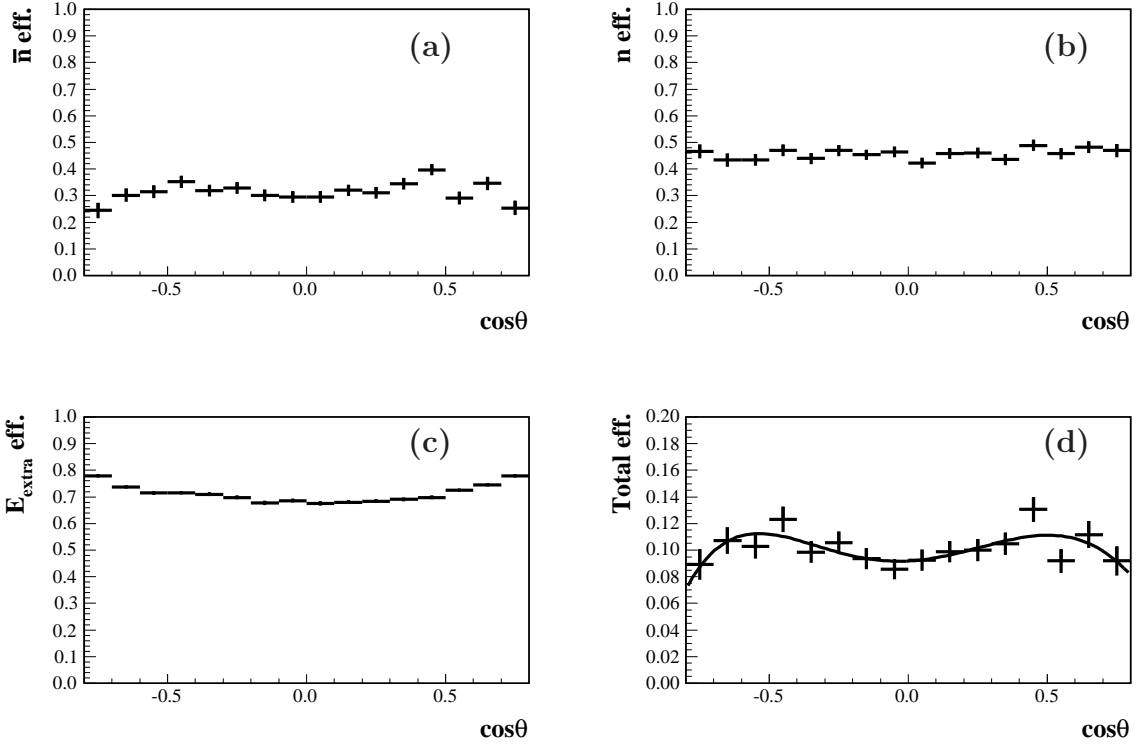}
\put(-300,270){\bf \large~(a)}
\put(-60,270){\bf \large~(b)}
\put(-300,120){\bf \large~(c)}
\put(-60,120){\bf \large~(d)}
\caption{Component selection efficiencies for $J/\psi\to n\bar{n}$ as a function 
of $\cos\theta$: (a)~$\bar{n}$ selection, (b)~$n$ selection, and (c)~$E_{extra}$ cut; 
(d)~overall product efficiency as computed (points) and smoothed by fitting to a 
fifth-order polynomial (line).} \label{fig-effnnbar}

\end{center}
\end{figure}

\newpage
\subsection{\bf Angular Distribution and branching fraction}

The number of $J/\psi\to n\bar{n}$ events in each $\cos\theta$ bin is obtained by fitting the distribution of the 
angle between the $n$ and $\bar{n}$.  The signal shape is determined with the $J/\psi\to p\bar{p}$ 
sample. Because of the 1.0-T magnetic field in the BESIII detector, the angular distribution in $J/\psi\to p\bar{p}$ must be corrected before being applied to $J/\psi\to n\bar{n}$. The signal shape ($\Psi$) for fitting the distribution of the angle between $n$ and $\bar{n}$ can be expressed in terms of 
$d\theta_{p}$ ($d\theta_{\bar{p}}$) and $d\phi_{p}$ ($d\phi_{\bar{p}}$), the polar and azimuthal 
angles between the shower position and the extrapolated EMC position of $p$ ($\bar{p}$), as follows:

$$\Psi=\sqrt{(1-\cos^2\theta_{p})d\phi_{p\bar{p}}^2+d\theta_{p\bar{p}}^2},$$

\noindent where $\theta_{p}$ is the polar angle of the proton track, 
$d\theta_{p\bar{p}}=d\theta_p+d\theta_{\bar{p}}$ and $d\phi_{p\bar{p}}=d\phi_p+d\phi_{\bar{p}}$. 
The background shape is fixed to the shape of the inclusive background, while signal and background 
normalizations are allowed to float in each $\cos\theta$ bin.  A sample fit for one $\cos\theta$ 
bin is shown in Fig.~\ref{fig-nnbardatacosfit}. 

After obtaining the bin-by-bin signal yields, we fit the resulting $\cos \theta$ distribution with 
the function $A(1+\alpha\cos^2\theta)\epsilon(\cos\theta)$, where $A$ gives the overall normalization 
and $\epsilon(\cos\theta)$ is the corrected efficiency. The resulting angular distribution and fit are 
shown in Fig.~\ref{fig-fitanglennbar}. The $\chi^2$ for the fit is 13 for 14 degrees of freedom, and 
the value determined for the angular-distribution parameter is $\alpha = 0.50\pm0.04$ (statistical 
error only).

The raw number of $J/\psi\to n\bar{n}$ events in the range $\cos\theta=[-0.8,0.8]$ is 
$N(-0.8,0.8) = 35891 \pm211$.  The efficiency-corrected yield obtained from the $\cos \theta$ fit is 
$N_{cor}(-0.8,0.8) = 354195 \pm2078$. The fitted value of $\alpha$ is used to determine the total 
number of $J/\psi \to n {\bar n}$ events in the full angular range of $\cos\theta=[-1.0,1.0]$
as follows:

$$N_{cor}(-1.0,1.0) = N_{cor}(-0.8,0.8) \times \frac{\int_{-1.0}^{1.0}(1+\alpha\cos^2\theta)d\theta}{\int_{-0.8}^{0.8}(1+\alpha\cos^2\theta)d\theta}=466590 \pm2737.$$

Combining this total yield with the number of $J/\psi$ events in our sample, we find the branching fraction to be

$$\mathcal{B}(J/\psi\to n\bar{n})=(2.07\pm0.01)\times10^{-3},$$

\noindent where the error is only statistical.

\begin{figure}[htbp]
\begin{center}
\includegraphics[width=8cm]{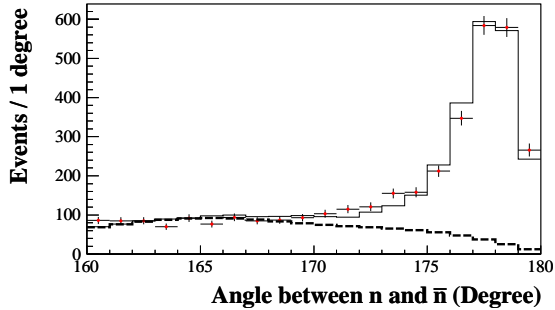}
\caption{Fit to the angle between the $n$ and $\bar{n}$ for $\cos\theta$ in [-0.3,-0.2].} \label{fig-nnbardatacosfit}
\end{center}
\end{figure}

\begin{figure}[htbp]
\begin{center}
\includegraphics[width=8cm]{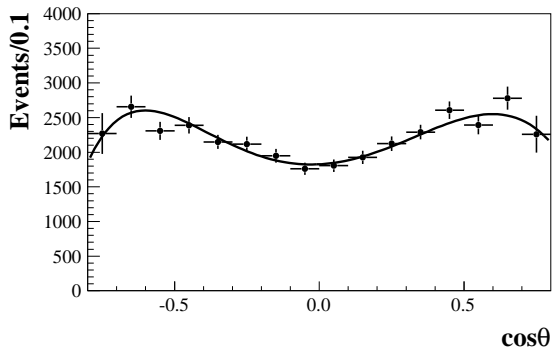}
\caption{The points represent the measured distribution of $\cos \theta$ for the $\bar{n}$ in 
$J/\psi\to n\bar{n}$ candidate events, with error bars that are the quadratic sums of the statistical 
and efficiency uncertainties. The line represents the fit of the distribution to the functional form 
given in the text, and is used to determine the normalization and the angular distribution parameter 
$\alpha$.} \label{fig-fitanglennbar}
\end{center}
\end{figure}

\subsection{\bf Systematic Errors and Results}


The different $\bar{n}~(n)$ momentum distributions in $J/\psi\to p\bar{n}\pi^-$ (c.c.) 
and $J/\psi\to n\bar{n}$ may introduce systematic uncertainties in the $\bar{n}~(n)$ efficiency determination. 
We change the missing-momentum range from ($1.1-1.2$)~GeV/$c$ to ($1.0-1.1$)~GeV/$c$ when selecting the $\bar{n}~(n)$ sample from $J/\psi\to p\bar{n}\pi^-$ (c.c.) and take the resulting differences  in $\alpha$ and the branching fraction as systematic errors. This estimation is cross-checked with a lower-statistics sample obtained from a separate BESIII data sample collected at the $\psi(3686)$ resonance. High-momentum $\bar{n}~(n)$ candidates are selected from the decay $\psi(3686) \to\pi^+\pi^- J/\psi, J/\psi\to p \bar{n} \pi^-$ (c.c.). The average $\bar{n}~(n)$ efficiencies obtained with this sample are consistent with those from $J/\psi\to p\bar{n} \pi^-$ (c.c.) within statistical errors.

A second source of systematic error in the $\bar{n}$ ($n$) efficiency is the effect of the requirement that the shower be within $10^\circ$ of the expected direction. We estimate the systematic errors due to this requirement by removing it 
and determining the changes in the results.  We sum these two systematic error in quadrature to 
obtain the total systematic errors due to the selection of $\bar{n}$ ($n$).  For the $\alpha$ 
determination, the $\bar{n}$ and $n$ errors are 0.04 and 0.09, respectively, and for the branching 
fraction they are $5 \times 10^{-5}$ and $1.2 \times 10^{-4}$.  For the branching fraction the $n$ 
efficiency is the largest source of uncertainty in our measurement.

As we did for $J/\psi\to p\bar{p}$, we use a toy MC method to estimate systematic errors due to 
the statistical uncertainties in the efficiency. For $\alpha$, this systematic error is $0.17$, 
the largest contributor to the overall uncertainty of the measurement.  For the branching 
fraction this error it is $7.1\times10^{-5}$ and is the second-largest contributor. 

We change the background shape for each exclusive MC background channel and repeat the fit of 
the angle between $n$ and $\bar{n}$. The largest variation observed for any case considered is assigned as the systematic error.

Our signal shape in fitting the $n-\bar{n}$ angle was obtained from $J/\psi\to p\bar{p}$. The correction of the $p-\bar{p}$ angular distribution into one appropriate for $n-\bar{n}$ in $J/\psi\to n\bar{n}$ is a source of systematic uncertainty.
To assess this we used a sideband subtraction instead of the fit to the angular distribution. We normalize the yield of MC background in the signal 
region ($170^{\circ}-180^{\circ}$)  by the numbers of events in the sideband range 
($160^{\circ}-170^{\circ}$)  for data and MC background. Then we take the background-subtracted 
number of events in the signal region ($170^{\circ}-180^{\circ}$) as the yield in each $\cos\theta$ 
bin. The differences between this alternative method and the standard method are assigned as 
systematic errors.  

The angular resolution can introduce systematic uncertainty both through the binning and through 
the $|\cos\theta|<0.8$ cut.  We use the $J/\psi\to p\bar{p}$ sample to evaluate the $\cos\theta$ 
resolution for $J/\psi\to n\bar{n}$.  In the data $d\cos\theta$ = 
$\cos\theta_{\bar{p}ext}-\cos\theta_{\bar{p}emc}$ is calculated as the equivalent of the 
resolution in $\cos\theta$ for $\bar{n}$, where $\cos\theta_{\bar{p}ext}$ represents the 
extrapolated position of the $\bar{p}$ at the EMC and $\cos\theta_{\bar{p}emc}$ is the 
reconstructed position in the EMC.  Here we assume that the position reconstruction of 
$\bar{n}$ in the EMC is similar to that for $\bar{p}$, because both are dominated by 
hadronic interactions.  To estimate the systematic error, we smear the $\cos\theta$ of 
$\bar{n}$ with the distribution of $d\cos\theta$ and redo the fit in each bin, and the fit to 
the angular distribution.  The resulting changes are taken as the systematic errors.  

For the all-neutral $n\bar{n}$ final state the trigger efficiency is another potential source 
of uncertainty. We correct the efficiency curve with the MC-determined trigger efficiency and  redo the 
fit.  The resulting changes in the branching fraction and $\alpha$ are taken as systematic 
uncertainties due to trigger efficiency.

To consider the systematic error from interference between the $J/\psi$ peak and the continuum, we write the total cross section $\sigma_{n\bar{n}}$ as 

$$\sigma_{n\bar{n}} = |\sqrt{\sigma_{n\bar{n}}^{cont.}} + \frac{\sqrt{12\pi\Gamma_{ee}\Gamma_{tot}}}{s-m^2+im\Gamma_{tot}}(E_{n}- e^{i\phi}S)|^2,$$ 
\\
where $E_{n}$ and $S$ are the EM and strong amplitudes for $J/\psi\to n\bar{n}$ and $\phi$ is the phase angle between them. $\sigma_{n\bar{n}}^{cont.}$ is the $n\bar{n}$ cross section contributed by continuum under the $J/\psi$ peak. These values are taken from Section~\ref{sec-summary} and, as discussed there, the EM amplitude $E_n$ should be opposite to $E_p$. The cross section $\sigma (e^+ e^- \rightarrow  n \bar{n})$ close to $J/\psi$  is assumed to lie between $\sigma (e^+ e^- \rightarrow  p \bar{p})$ and $\sigma (e^+ e^- \rightarrow  p \bar{p}) \cdot (\mu_n /\mu_p)^2$, so the $E_n$, which is in proportion to $\sqrt{\sigma_{n\bar{n}}^{cont.}}$, ranges from $E_p$ to $E_p(\mu_n /\mu_p)$, where $\mu_n$ and $\mu_p$ are the magnetic moments of the neutron and proton. To estimate the uncertainty from the continuum, we take the larger one, $\sigma_{n\bar{n}}^{cont.} \sim \sigma_{p\bar{p}}^{cont.}$, therefore also $E_{n} \sim  E_{p}$. The difference with and without $\sigma_{n\bar{n}}^{cont.}$ is assigned as systematic error.

Finally, we change $\alpha$  by $\pm 1$~$\sigma$ (including the systematic error) and reevaluate the branching fraction to estimate the systematic error in the branching fraction 
measurement. 

Table~\ref{tab-nnbarsys}  summarizes the systematic uncertainties and their sum in quadrature.  
The final results for our $J/\psi\to n\bar{n}$ 
measurements are as follows:

$$\alpha=0.50\pm0.04\pm0.21,~{\rm and}$$

$$\mathcal{B}(J/\psi\to n\bar{n})=(2.07\pm0.01\pm0.17)\times10^{-3}.$$

\noindent The branching fraction measurement is consistent with the previous world 
average~\cite{ref:pdg} and improves the overall precision by about a factor of 2.3.

\begin{table*}[htbp]
\begin{center}
\caption{Systematic errors for $J/\psi\to n\bar{n}$.} \label{tab-nnbarsys}
\vspace{0.2cm}{
\begin{tabular}{|c|c|c|}\hline\hline
Sources & Effect on $\alpha$ & Effect on $\mathcal{B}$ ($10^{-3}$)\\
\hline\hline
$\bar{n}$ selection  & 0.04 & 0.05\\
$n$ selection  & 0.09 & 0.12\\
Efficiency correction statistics & 0.17 & 0.07\\
Background & 0.03 & 0.03\\
Signal shape & 0.02 & 0.06\\
$\cos\theta$ resolution & 0.05 & 0.01\\
Trigger & 0.03 & 0.01\\
$\alpha$ value & $-$ & 0.01\\
number of $J/\psi$ & $-$ & 0.03\\
continuum & $-$ & 0.01\\
\hline
Total & 0.21 & 0.17 \\
\hline\hline
\end{tabular}
}
\end{center}
\end{table*}

\section{\bf Summary}\label{sec-summary}

We have used the world's largest sample of $J/\psi$ decays to make new measurements of the 
branching fractions and production-angle distributions for $J/\psi\to p\bar{p}$ and 
$J/\psi\to n\bar{n}$, obtaining the branching fractions 
$\mathcal{B}(J/\psi\to p\bar{p})=(2.112\pm0.004\pm0.031)\times10^{-3}$
and $\mathcal{B}(J/\psi\to n\bar{n})=(2.07\pm0.01\pm0.17)\times10^{-3}$.  These results 
represent significant improvements over previous measurements. The angular distributions for both decays are well described 
by the functional form $1+\alpha\cos^2\theta$, with measured values of 
$\alpha=0.595\pm0.012\pm0.015$ for $J/\psi\to p\bar{p}$, and 
$\alpha=0.50\pm0.04\pm0.21$ for $J/\psi\to n\bar{n}$.

The  $p \bar{p}$ angular distribution can be decomposed as $|C^M_p|^2 (1+\cos^2\theta)+(2M_p/M_{J/\psi})^ 2 |C^E_p|^2 \sin^2\theta$,  where $C^M_p$ and $C^E_p$ are the total helicity $\pm 1$ and 0 decay amplitudes. In terms of the angular parameter $\alpha$, the ratio of amplitudes is $|C^E_p/C^M_p| = M_{J/\psi} / (2 M_p)  \sqrt{(1-\alpha)/(1+\alpha)}$. With our measured values for $\alpha$, we find 
$|C^E_p/C^M_p|= 0.832 \pm 0.015 \pm 0.019$ and $|C^E_n/C^M_n|  = 0.95 \pm 0.05 \pm 0.27$, respectively. These measurements permit discrimination among the different proposed models 
\cite{angletheory1, angletheory2, angletheory3, angletheory4, angletheory5, angletheory6, angletheory7}.

The relative phase between the strong and EM amplitudes can be obtained by comparing $\mathcal{B}(J/\psi \rightarrow p \bar{p})$ and $\mathcal{B}(J/\psi \rightarrow n \bar{n})$. The $J/\psi$ EM decay amplitudes are related to the corresponding continuum cross sections close to the $J/\psi$ as follows: $E_N^2(J/\psi  \rightarrow  \gamma^* \rightarrow  N \bar{N})$ = $\mathcal{B}(J/\psi \rightarrow  \mu \mu)  \cdot \sigma (e^+ e^- \rightarrow  N \bar{N}) / \sigma(e^+ e^- \rightarrow \mu \mu)$. Present data~\cite{Aubert:2005cb} suggest that $\sigma (e^+ e^- \rightarrow  p \bar{p})  \sim  (9 \pm 3)~\rm{pb}$,  if fitted with a smooth $ W^{-10}$  as expected at high enough center-of-mass energies $W$.  For  $\sigma (e^+ e^- \rightarrow  n \bar{n})$  the only available data \cite{Antonelli:1998fv,hadron2011}  are close to threshold. In the following it is assumed that the neutron time-like dominant  magnetic form factor  is negative at these center-of-mass energies, like the magnetic moment, as predicted by dispersion relations~\cite{Baldini:1998qn}. The cross section $\sigma (e^+ e^- \rightarrow  n \bar{n})$ close to $J/\psi$  is assumed to lie between $\sigma (e^+ e^- \rightarrow  p \bar{p})$, as is seen in the present data close to threshold, and $\sigma (e^+ e^- \rightarrow  p \bar{p}) \cdot (\mu_n /\mu_p)^2$, as in the space-like region~\cite{qcd2009}.
Taking into account these hypotheses and their overall uncertainties, and neglecting the contribution of continuum amplitudes, the strong amplitude $S$ is given by
$$S^2 = [(\mathcal{B}(J/\psi\to p\bar{p})-E^2_p)E_n+ (\mathcal{B}(J/\psi\to n\bar{n})-E^2_n)E_p] / (E_n+E_p) = (2.038 \pm  0.094 ) \cdot 10^{-3},$$
 and the phase $\phi$ between the strong and EM amplitudes is found to be
$$ \phi = \cos^{-1}[ (\mathcal{B}(J/\psi\to p\bar{p})-S^2-E^2_p)/(2 SE_p)] = (88.7  \pm   8.1)^\circ.$$ 
The uncertainty in the phase is mostly due to the $\mathcal{B}(J/\psi \rightarrow n \bar{n})$ systematic error. This determination confirms the orthogonality of the strong and EM amplitudes within the precision of our measurement.

\section{\bf ACKNOWLEDGMENTS}
The BESIII collaboration thanks the staff of BEPCII and the computing center for their hard efforts. This work is supported in part by the Ministry of Science and Technology of China under Contract No. 2009CB825200; Joint Funds of the National Natural Science Foundation of China under Contracts Nos. 11079008, 11179007; National Natural Science Foundation of China (NSFC) under Contracts Nos. 10625524, 10821063, 10825524, 10835001, 10935007, 11125525; the Chinese Academy of Sciences (CAS) Large-Scale Scientific Facility Program; CAS under Contracts Nos. KJCX2-YW-N29, KJCX2-YW-N45; 100 Talents Program of CAS; Istituto Nazionale di Fisica Nucleare, Italy; U. S. Department of Energy under Contracts Nos. DE-FG02-04ER41291, DE-FG02-91ER40682, DE-FG02-94ER40823, DE-FG02-05ER41374; U.S. National Science Foundation; University of Groningen (RuG) and the Helmholtzzentrum fuer Schwerionenforschung GmbH (GSI), Darmstadt; WCU Program of National Research Foundation of Korea under Contract No. R32-2008-000-10155-0

\end{document}